\documentclass[pra, showpacs, twocolumn, floatfix]{revtex4}
\usepackage{graphicx}
\usepackage{epstopdf}
\usepackage{amsmath, amsfonts, amssymb, bm}
\begin{document}
\title{Quantum correlations among optical and vibrational quanta}
\author{Sergiu \surname{Carlig}}

\author{Mihai A. \surname{Macovei}}
\email{macovei@phys.asm.md}
\affiliation{Institute of Applied Physics, Academy of Sciences of Moldova, 
Academiei str. 5, MD-2028 Chi\c{s}in\u{a}u, Moldova}
\date{\today}
\begin{abstract}
We investigate the feasibility of correlating an optical cavity field and a vibrational phonon mode. A laser pumped quantum dot fixed on 
a nano-mechanical resonator beam interact as a whole with the optical resonator mode. When the quantum dot variables are faster than the 
optical and phonon ones, we obtain a final master equation describing the involved modes only. Increasing the temperature, that directly affects 
the vibrational degrees of freedom, one can as well influence the cavity photon intensity, i.e., the optical and phonon modes are correlated. 
Furthermore, the corresponding Cauchy-Schwarz inequality is violated demonstrating the quantum nature of those correlations.
\end{abstract}
\maketitle

\section{Introduction}
Nano-mechanical resonators are actively and widely investigated \cite{prm,rew_nm}. Their importance for ultra-light mass detection, ultra-small displacements 
or very small force measurements is now well recognized \cite{rew_nm} (and references therein). Significant contributions were achieved 
towards macroscopic  quantum effects in these systems combined with existing ones. For instance, superconducting qubit storage and entanglement 
with nano-mechanical resonators was investigated in \cite{qb1} while feedback-enhanced parametric squeezing of mechanical motion was investigated in 
Ref.~\cite{sq1}, respectively. Observation of quantum motion of a nano-mechanical resonator was reported in \cite{exp_m}. Coherent phonon 
manipulation in coupled mechanical resonators was experimentally demonstrated in \cite{man1} and this may help in entangling two distinct macroscopic 
mechanical objects \cite{ent1}. A nano-mechanical interface between optical photons and microwave electrical signals was recently demonstrated as well 
\cite{inter}. Furthermore, the dipole-dipole coupling among a small ensemble of Rydberg atoms allows for quantum control of macroscopic mechanical 
oscillators via their mutual interactions \cite{mey} whereas stationary continuous-variable entanglement between an optical cavity and a nano-mechanical 
resonator beam was demonstrated too, in Ref.~\cite{ent2}. Finally, phonon lasing was experimentally achieved in an electromechanical resonator \cite{las1}.

Here, we investigate a system composed of a two-level quantum dot fixed on a suspended nano-mechanical resonator inside an optical cavity.
The quantum dot is externally laser pumped. Spontaneous decay as well as optical or vibrational modes damping rates are correspondingly taking 
into account. We are interested in a regime where the quantum dot dynamics is faster than the dynamics of other subsystems involved, i.e. in the 
good cavity limit. In this way, one can reduce the whole system to the quantum dynamics of the two modes, vibrational and optical. We found 
quantum correlations among the optical and vibrational quanta under particular circumstances of stronger qubit-resonator coupling strengths. Actually, 
we have demonstrated the violation of the Cauchy-Schwarz inequality defined as a product of the photon-photon and phonon-phonon second-order 
correlation functions divided on the photon-phonon or phonon-photon cross-correlation function squared. Moreover, the environmental temperature that 
explicitly affects the mechanical subsystem only modifies as well the mean-photon number in the optical resonator mode demonstrating the existence of 
correlations among the photon and phonon modes. 

The article is organized as follows. In Sec. II we describe the model as well as the analytical approach used. The master equation characterizing the 
induced correlations among optical and phonon modes is derived there. Sec. III deals with the corresponding equations of motion and discussion of 
the obtained results. The Summary is given in the last section, i.e. Sec. IV.

\section{Analytical approach}
The investigated model is described as follows: A two-level quantum dot of frequency $\omega_{0}$ is fixed on a semiconductor 
beam structure inside an optical cavity.  A coherent laser field with the wave-vector $\vec k_{L}$ is resonantly interacting with the two-level quantum 
dot leading to correlations between mechanical vibrations and photon scattering. If the thickness of the beam is smaller than its width, the lowest-energy 
resonance corresponds to the fundamental flexural mode with the frequency being of the order of $\rm{GHz}$.  Flexions induce extensions and compressions 
in the structure modifying the deformation potential coupling of the embedded quantum dot and, hence, its energy levels \cite{prm1}. Concomitantly, the 
artificial two-level emitter interacts with the optical cavity mode. The vibrational and optical mode frequencies are denoted by $\omega$ and $\omega_{c}$, 
respectively. The model Hamiltonian describing the whole system is:
\begin{eqnarray}
H &=& \hbar\omega_{c}a^{\dagger}a + \hbar\omega b^{\dagger}b + \hbar\omega_{0}S_{z} 
+ \hbar g(a^{\dagger}S^{-} +  aS^{+}) \nonumber \\
&+& \hbar \Omega(S^{+}e^{-i\omega_{L}t} + S^{-}e^{i\omega_{L}t}) + \hbar\lambda S_{z} (b^{\dagger} + b). 
\label{Hm}
\end{eqnarray}
Here, the first three terms describe the free energies of optical and mechanical modes as well as of the artificial two-level system. The fourth and the fifth 
terms characterize the interaction of the quantum dot with the optical resonator mode and laser field, respectively. The last term takes into account the 
interaction of the vibrational degrees of freedom with the radiator \cite{prm1}. $g$ and $\lambda$ denote the interaction strengths among the two-level 
emitter and the involved optical and mechanical modes, while $\Omega$ is the corresponding Rabi frequency due to external laser pumping. The qubit 
operators $S_{z}$ and $S^{\pm}$ have the usual meaning and satisfy the standard commutation relations. $\{a^{\dagger}, b^{\dagger}\}$ 
and $\{a,b\}$ are the generation and annihilation operators for photon and phonon subsystems, respectively, and obey the boson commutation 
relations \cite{kmek,book}.

The entire system will be described in the dressed-state representation \cite{book,DM}: $|g\rangle=\sin{\theta}|+\rangle + \cos{\theta}|-\rangle$ and 
$|e\rangle=\cos{\theta}|+\rangle - \sin{\theta}|-\rangle$, where $\cot{2\theta}=\Delta/2\Omega$  with $\Delta=\omega_{0}-\omega_{L}$ being 
the detuning of the laser frequency $\omega_{L}$ from the two-level transition frequency. $|e\rangle$ and $|g\rangle$ are the excited and the 
ground bare states of the quantum dot, while $|+\rangle$ and $|-\rangle$ the corresponding states in the dressed-state picture. In the interaction 
picture, the master equation describing our model in the rotating-wave and dipole approximations as well as in Born-Markov approximations is:
\begin{eqnarray}
\dot{\rho}&+&\frac{i}{\hbar}[H_{d},\rho]=-\gamma_{0}[R_{z},R_{z}\rho]-\gamma_{+} [R_{\pm},R_{\mp}\rho]\nonumber\\
&-&\gamma_{-}[R_{\mp},R_{\pm}\rho] - \kappa_{a}[a^\dagger,a\rho] - \kappa_{b}(1+\bar{n})[b^\dagger,b\rho]\nonumber\\
&-&\kappa_{b}\bar{n}[b,b^\dagger\rho] + H.c.,\label{me_pm}
\end{eqnarray}
where an overdot denotes differentiation with respect to time. 
Here, the dressed-state Hamiltonian is $ H_{d}=H_{0}+H_{i}$ with:
\begin{eqnarray}
H_{0}&=&\hbar\Omega_R R_z-\hbar\Delta_{1}a^{\dagger}a + \hbar\omega b^{\dagger}b, \nonumber \\
H_{i}&=&{A} R_z+{B}^{\dagger}R_{\mp}e^{-2i\Omega_{R}t}+{B}R_{\pm}e^{2i\Omega_{R}t}, \label{H0I} 
\end{eqnarray}
where $\Delta_{1}=\omega_{L} - \omega_{c}$ and $\Omega_{R} = \sqrt{(\Delta/2)^2 + \Omega^2}$, whereas
\begin{eqnarray}
{A}&=&\frac{\hbar}{2}\bigl( g(ae^{i\Delta_{1}t}+a^{\dagger}e^{-i\Delta_{1} t})\sin{2\theta}\nonumber\\
&+&\lambda(be^{-i\omega t}+b^{\dagger}e^{i\omega t})\cos{2\theta}\bigr), \nonumber\\
{B}&=&\hbar g(ae^{i\Delta_{1}t}\cos^{2}{\theta}-a^{\dagger}e^{-i\Delta_{1}t}\sin^{2}{\theta})\nonumber\\
&-&\frac{\hbar}{2}\lambda \sin{2 \theta}(be^{-i\omega t}+b^{\dagger}e^{i\omega t}).
\end{eqnarray}
The dressed-state quantum dot operators are defined as follows: $R_{z}=|+\rangle \langle+| - |-\rangle \langle-|$, $R_{\pm}=|+\rangle \langle-|$ and 
$R_{\mp}=|-\rangle \langle+|$  satisfying the standard commutation relations for su(2) algebra. Further, $\gamma_{0}$=$\frac{1}{4}(\gamma \sin^2{2\theta}
+\gamma_{c}\cos^2{2\theta})$, $\gamma_{+}$=$\gamma \cos^4{\theta} + \frac{\gamma_{c}}{4}\sin^{2}{2\theta}$ and $\gamma_{-}=\gamma \sin^4{\theta}
+ \frac{\gamma_{c}}{4}\sin^{2}{2\theta}$ describe the spontaneous decay processes among the involved dressed-states, while $\gamma$ and $\gamma_c$ 
are the single-qubit spontaneous decay and dephasing rates, respectively. $\kappa_{a}(\kappa_{b})$ is the photon (phonon) resonator damping rate whereas 
$\bar n$ is the mean-phonon number corresponding to vibrational frequency $\omega$ and environmental temperature $T$.

The master equation (\ref{me_pm}) is quite complex. However, for our purposes one can significantly simplify it. 
Particularly, we are interested in a regime where the pumped quantum-dot system is faster than the cavity photon 
and vibrational phonon subsystems, respectively. Consequently, the quantum dot variables can be eliminated from 
the whole quantum dynamics, an approximation valid for $\Omega \gg \gamma \gg \kappa_{a,b}$ as well as $\Omega \gg \{ g, \lambda \}$. 
Notice that this approach is widely used in other somehow related systems \cite{zb1,sc_zb,chk,gxl}. In the following, we write down the master equation (\ref{me_pm}) 
after tracing over quantum-dot degrees of freedom:
\begin{eqnarray}
\dot{\rho}_f&=&-i[{A},\rho_{++}-\rho_{--}]-i[{B}^\dagger,\rho_{+-}]e^{-2i\Omega_R t}\nonumber\\
&-&i[{B},\rho_{-+}]e^{2i\Omega_R t} - \kappa_{a}[a^\dagger,a\rho_{f}]\nonumber\\
&-&\kappa_{b}(1+\bar{n})[b^\dagger,b\rho_{f}] - \kappa_{b}\bar{n}[b,b^\dagger\rho_{f}] + H.c., \label{rf}
\end{eqnarray}
where $\rho_{\alpha\beta}=\langle \alpha|\rho_{q}|\beta\rangle \rho_{f}$, with $\{\alpha,\beta \in \pm \}$. The quantum dot variables can be found 
from Eq.~(\ref{me_pm}), namely:
\begin{eqnarray}
\dot{\rho}_{+-}&=&-\Gamma_{\perp}\rho_{+-}-i({A}\rho_{+-} + \rho_{+-}{A})\nonumber\\
&-&i({B}\rho_{--}-\rho_{++}{B})e^{2i\Omega_{R}t},\nonumber\\
\dot{\rho}_{++}&=&-2\gamma_{+}\rho_{++}+2\gamma_{-}\rho_{--}-i({A}\rho_{++}-\rho_{++}{A})\nonumber\\
&-&i({B}\rho_{-+}e^{2i\Omega_R t}-\rho_{+-}{B}^{\dagger}e^{-2i\Omega_{R}t}),\nonumber\\
\dot{\rho}_{--}&=&2\gamma_{+}\rho_{++} -2\gamma_{-}\rho_{--} + i({A}\rho_{--} - \rho_{--}{A})\nonumber\\
&+&i(\rho_{-+}{B} e^{2i\Omega_{R}t} - {B}^{\dagger}\rho_{+-}e^{-2i\Omega_{R}t})\label{rho+-}, \label{eks}
\end{eqnarray}
with $\Gamma_{\perp}=4\gamma_0+\gamma_{+}+\gamma_{-}$.
In the secular approximation and to first order in the interaction parameters $\{g,\lambda\}$ the solutions of  (\ref{eks}) are:
\begin{eqnarray}
&&\rho_{+-}=-i e^{2i\Omega_R t}(\bar{B}\rho_{--}-\rho_{++}\bar{B}), \nonumber\\
&&\rho_{++}-\rho_{--}=- i(\bar{A}\rho_f-\rho_f\bar{A}), \label{esol}
\end{eqnarray}
where
\begin{eqnarray*}
\bar{A}&=&\frac{g}{2}\biggl(\frac{a^\dagger e^{-i\Delta_1 t}}{\Gamma_{\shortparallel} -i\Delta_{1}}+\frac{a e^{i\Delta_1 t}}
{\Gamma_{\shortparallel}+i\Delta_{1}} \biggr)\sin{2\theta}\nonumber\\
&+&\frac{\lambda}{2}\biggl(\frac{b^\dagger e^{i\omega t}}{\Gamma_{\shortparallel} +i\omega}+\frac{b e^{-i\omega t}}
{\Gamma_{\shortparallel}-i\omega}\biggr)\cos{2\theta},\nonumber\\
\bar{B}&=&-\frac{\lambda}{2}\biggl(\frac{\sin{2\theta}b^\dagger e^{i\omega t}}{\Gamma_{\perp}+i(2 \Omega_{R} + \omega)} + \frac{\sin{2\theta}b e^{-i\omega t}}{\Gamma_{\perp} + i(2 \Omega_{R} - \omega)}\biggr)\nonumber\\
&+&g\biggl( \frac{a \cos^2{\theta} e^{i\Delta_1 t}}{\Gamma_{\perp}+i(2 \Omega_R+\Delta_1)}-\frac{a^\dagger\sin^2{\theta} e^{-i\Delta_1 t}}{\Gamma_{\perp}+i(2 \Omega_R-\Delta_1)} \biggr). 
\end{eqnarray*}
Here $\Gamma_{\shortparallel} = \gamma(1 + \cos^{2}{2\theta})+\gamma_{c}\sin^2{2\theta}$. Inserting the solutions given by (\ref{esol}) in Eq.~(\ref{rf})
and using the identity ${\rm Tr}\{\dot{\rho}(t)Q\}={\rm Tr}\{\dot Q(t)\rho\}$, one can obtain the final master equation describing the quantum dynamics of the 
photon and phonon subsystems in the good cavity limit:
\begin{eqnarray}
&{}&\langle\dot{Q}\rangle + \frac{i}{2}(\Delta_1-\omega)\langle[a^\dagger a+b^\dagger b,Q]\rangle=\nonumber\\
&=&\langle[Q,a](-A^{*}_{1}a^{\dagger}+D^{*}_{2}b)+(B^{*}_{1}a^{\dagger}-C^{*}_{2}b)[Q,a]\rangle\nonumber\\
&+&\langle[Q,a^{\dagger}](-B_{1}a+C_{2}b^{\dagger})+(A_{1}a-D_{2}b^{\dagger})[Q,a^{\dagger}]\rangle\nonumber\\
&+&\langle[Q,b](D^{*}_{1}a-A^{*}_{2}b^{\dagger})+(B^{*}_{2}b^{\dagger}-C^{*}_{1}a)[Q,b]\rangle\nonumber\\
&+&\langle[Q,b^{\dagger}](C_{1}a^{\dagger}-B_{2}b)+(A_{2}b - D_{1}a^{\dagger})[Q,b^{\dagger}]\rangle \label{Q}. \label{Qq}
\end{eqnarray}
Here $Q$ denotes, respectively, any operator belonging to the photon and phonon subsystems, while
\begin{eqnarray*}
A_{1}&=&\frac{1}{4}\frac{g^2 \sin ^2{2 \theta}}{\Gamma_{\shortparallel} +i \Delta_{1} }+\frac{g^2 P_{-} \sin
   ^4{\theta}}{\Gamma_{\perp} -i (2 \Omega_{R} -\Delta_{1} )}\nonumber\\
&+&\frac{g^2 P_{+} \cos ^4{\theta}}{\Gamma_{\perp} + i (2 \Omega_{R}+\Delta_{1}  )}, \nonumber\\
A_{2}&=&\frac{1}{4}\bigg(\frac{\lambda^2 \cos^{2}{2 \theta}}{\Gamma_{\shortparallel} -i \omega }+\frac{\lambda^2 P_{-}
\sin^2{2\theta}}{\Gamma_{\perp} -i (2 \Omega_{R} +\omega )}\nonumber\\
&+&\frac{\lambda^2 P_{+} \sin ^2{2 \theta}}{\Gamma_{\perp} + i(2 \Omega_{R}-\omega  )}\bigg)+\kappa_{b}\bar{n}, \nonumber\\
C_{1}&=&\frac{P_{+}}{2}\frac{g\lambda \sin{2 \theta}\cos^2{\theta}}{\Gamma_{\perp} -i (2\Omega_{R}+\Delta_{1}) }
-\frac{P_{-}}{2}\frac{g\lambda \sin{2 \theta}\sin^2{\theta}}{\Gamma_{\perp}+ i (2\Omega_{R}-\Delta_{1})} \nonumber\\
&-&\frac{1}{4}\frac{g\lambda \sin{2 \theta}\cos{2\theta}}{\Gamma_{\shortparallel} -i\Delta_{1}}, \nonumber\\
C_{2}&=&\frac{P_{-}}{2}\frac{g\lambda \sin{2 \theta}\cos^2{\theta}}{\Gamma_{\perp} +i (2\Omega_{R}+\omega) }
-\frac{P_{+}}{2}\frac{g\lambda \sin{2 \theta}\sin^2{\theta}}{\Gamma_{\perp}- i (2\Omega_{R}-\omega)}\nonumber\\
&-&\frac{1}{4}\frac{g\lambda \sin{2 \theta}\cos{2\theta}}{\Gamma_{\shortparallel} + i\omega },
\end{eqnarray*}
with
\begin{eqnarray*}
P_{+} = \frac{\gamma_-}{\gamma_++\gamma_-}, ~~{\rm and}~~ P_{-}=\frac{\gamma_+}{\gamma_++\gamma_-}.
\end{eqnarray*}
$B_{i}$ can be obtained from $A_{i}$ via $P_{\mp} \leftrightarrow P_{\pm}$ as well as by adding $\kappa_{a}$ to $B_{1}$ and $\kappa_{b}$ to $B_{2}$, 
correspondingly. Respectively, $D_{i}$ can be obtained from $C_{i}$ through $P_{\mp} \leftrightarrow P_{\pm}$, and $\{i \in 1,2\}$. Notice that in 
obtaining Eq.~(\ref{Qq}) we have ignored rapidly oscillating terms at frequencies: $\pm 2\Delta_{1}, \pm (\Delta_{1}+ \omega)$ and $\pm 2\omega$.
Actually, we are interesting in a regime where a laser photon absorption is accompanied by the generation of a phonon and an optical cavity 
photon, respectively, that is when $\Delta_{1} \approx \omega$.

In the following section, we shall describe the induced quantum correlations among the optical and mechanical degrees of freedom.

\section{Quantum correlations}
Eq.~(\ref{Qq}) allows us to obtain the equations of motion for the variables of interest. We can define $Q=a^{\dagger j} a^{k} b^{\dagger l} b^{m}$ 
(where $\{j,k,l,m\}$ are any integer numbers) as a general operator belonging to the both, photon and phonon subsystems. The equation of motions  
for the mean-values of photon and phonon numbers etc. can be then obtained from the following main equation:
\begin{eqnarray}
&&\frac{d}{dt}\langle a^{\dagger j}  a^k  b^{\dagger l}  b^m \rangle=\langle a^{\dagger j}  a^k  b^{\dagger l}  b^m \rangle \nonumber\\
&\times&\bigr((A_{1}^*-B_{1}^*)j+(A_{1}-B_{1})k+(A_{2}^*-B_{2}^*)l\nonumber\\
&+&(A_{2}-B_{2})m-\frac{i}{2}(\Delta_1-\omega)(j-k+l-m)\big)\nonumber\\
&+&\langle a^{\dagger j+1}  a^k  b^{\dagger l}  b^{m-1}\rangle(C_{1}-D_{1})m\nonumber\\
&+&\langle a^{\dagger j-1}  a^k  b^{\dagger l}  b^{m+1}\rangle(C_{2}^*-D_{2}^*)j\nonumber\\
&+&\langle a^{\dagger j}  a^{k+1}  b^{\dagger l-1}  b^{m}\rangle(C_{1}^*-D_{1}^*)l\nonumber\\
&+&\langle a^{\dagger j}  a^{k-1}  b^{\dagger l+1}  b^{m}\rangle(C_{2}-D_{2})k\nonumber\\
&+&\langle a^{\dagger j-1}  a^k  b^{\dagger l-1}  b^{m}\rangle(C_{1}^*+C_{2}^*)jl\nonumber\\
&+&\langle a^{\dagger j-1}  a^{k-1}  b^{\dagger l}  b^{m}\rangle(A_{1}+A_{1}^*)jk\nonumber\\
&+&\langle a^{\dagger j}  a^{k-1}  b^{\dagger l}  b^{m-1}\rangle(C_{1}+C_{2})km\nonumber\\
&+&\langle a^{\dagger j}  a^k  b^{\dagger l-1}  b^{m-1}\rangle(A_{2}+A_{2}^*)lm. \label{aabb}
\end{eqnarray}
For instance, by selecting instead of $\{j,k,l,m\}$ the following sets: $\{1,1,0,0\},\ \{0,0,1,1\}, \{0,1,0,1\},\{1,0,1,0\}$ one arrives at the 
equations of motion for the mean-values of the photon and phonon numbers and their first-order correlations, namely:
\begin{eqnarray}
\frac{d}{dt}\langle a^{\dagger} a \rangle& = &\langle a^{\dagger} a\rangle(A_{1} - B_{1} + A_{1}^{*} - B_{1}^{*}) + \langle ab\rangle(C_{2}^{*} - D_{2}^{*})
\nonumber\\
&+&\langle a^{\dagger} b^{\dagger}\rangle (C_{2} - D_{2}) + A_{1} + A_{1}^{*}, \nonumber\\
\frac{d}{dt}\langle b^{\dagger} b\rangle& = &\langle b^{\dagger} b\rangle(A_{2} - B_{2} + A_{2}^{*} - B_{2}^{*}) + \langle ab\rangle(C_{1}^{*} - D_{1}^{*})\nonumber\\
&+&\langle a^{\dagger}b^{\dagger}\rangle (C_{1} - D_{1}) + A_{2} + A_{2}^{*},\nonumber\\
\frac{d}{dt}\langle ab\rangle &=& \langle ab\rangle \bigl(A_{1} - B_{1} + A_{2} - B_{2} + i(\Delta_{1} - \omega)\bigr)\nonumber\\
&+& \langle a^{\dagger}a\rangle(C_{1} - D_{1}) + \langle b^{\dagger}b\rangle (C_{2} - D_{2}) + C_{1} + C_{2}, \nonumber\\
\frac{d}{dt}\langle a^{\dagger}b^{\dagger}\rangle &=& \langle a^{\dagger} b^{\dagger}\rangle\bigl(A_{1}^{*} - B_{1}^{*} + A_{2}^{*} - B_{2}^{*} - 
i(\Delta_{1} - \omega)\bigr)\nonumber\\
&+&\langle a^{\dagger} a\rangle(C_{1}^{*} - D_{1}^{*}) + \langle b^{\dagger} b\rangle (C_{2}^{*} - D_{2}^{*}) + C_{1}^{*} + C_{2}^{*}. \label{aa}
\nonumber\\ \label{eksm}
\end{eqnarray}
Additionally, the equations of motion for the second-order correlation functions and their cross-correlations, that is:
\begin{eqnarray}
g_{1}^{(2)}(0)&=&\frac{\langle{a^{\dagger}  a^{\dagger} a a }\rangle}{\langle a^{\dagger}a\rangle^2}, ~~~
g_{2}^{(2)}(0)=\frac{\langle{b^{\dagger}  b^{\dagger} b b }\rangle}{\langle b^{\dagger}b\rangle^2}, \nonumber\\
g_{3}^{(2)}(0)&=&\frac{\langle{a^{\dagger}  a b^{\dagger} b }\rangle}{\langle a^{\dagger}a\rangle \langle b^{\dagger}b\rangle}, \label{gg}
\end{eqnarray}
can be obtained again from Eq.~(\ref{aabb}) by considering, respectively, the next sets of numbers: $\{2,2,0,0\}$, $\{0,0,2,2\}$, $\{1,1,1,1\}$, $\{1,0,2,1\}$, 
$\{0,1,1,2\}$, $\{2,1,1,0\}$, $\{1,2,0,1\}$, $\{2,0,2,0\}$, $\{0,2,0,2\}$ instead of $\{j,k,l,m\}$.  An efficient tool to investigate the quantum features of the 
correlations between photons and phonons  is the Cauchy-Schwarz inequality, CSI, which is defined as follows \cite{csi}:
\begin{eqnarray}
{\rm CSI} = g^{(2)}_{1}(0)g^{(2)}_{2}(0)/[g^{(2)}_{3}(0)g^{(2)}_{3}(0)]. \label{csi}
\end{eqnarray}
Particularly, the induced correlations are of quantum nature if ${\rm CSI} < 1$. 
\begin{figure}[t]
\centering
\includegraphics[height=5.3cm]{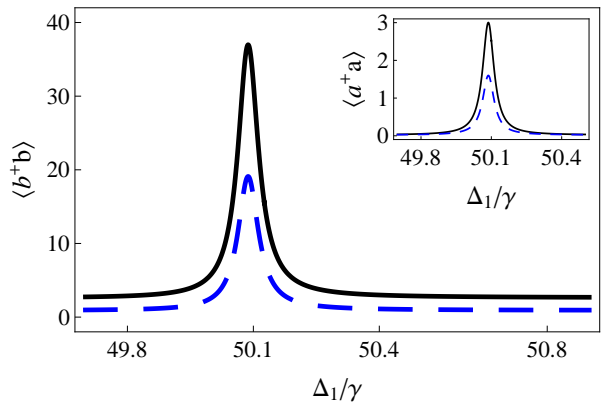}
\caption{\label{fig-1} 
(color online) The mean-value of the phonon number $\langle b^{\dagger}b\rangle$ as a function of $\Delta_{1}/\gamma$. 
Here, $\gamma_{c}/\gamma=0.3$, $g/\gamma=3$, $\lambda/\gamma=5$, $\Omega/\gamma=50$, $\omega/\gamma=50$, 
$\Delta/(2\Omega)=-0.263$, $\kappa_{a}/\gamma=0.09$ and $\kappa_{b}/\gamma=0.009$. The solid line corresponds to $\bar n=2$, 
whereas the dashed one to $\bar n=0.5$. The inset figure shows the same but for the photon mean-number $\langle a^{\dagger}a\rangle$, 
respectively.}
\end{figure}
\begin{figure}[t]
\centering
\includegraphics[height=5.4cm]{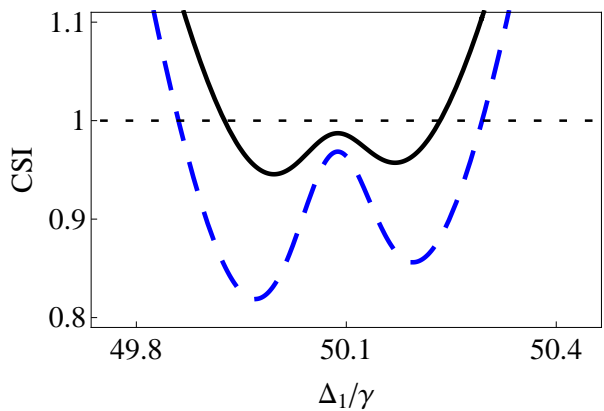}
\caption{\label{fig-2} 
(color online) The steady-state value of the CSI versus $\Delta_{1}/\gamma$. 
The solid line corresponds to $\bar n=2$, whereas the long-dashed dashed one to $\bar n=0.5$. 
Violation of the CSI occurs below the horizontal dashed line. Other parameters are the same as in Fig.~{(\ref{fig-1})}.}
\end{figure}

Figure (\ref{fig-1}) shows the steady-state value of the mean phonon and photon numbers for particular parameters close to those already 
considered in other setups \cite{prm,rew_nm,ent2,prm1,DM,prm2}. For stronger qubit-cavity couplings one can observe a peak of 
these quantities around $\Delta_{1} \approx \omega$, i.e.  where a laser photon absorption is accompanied by the generation of a phonon 
and an optical cavity photon, respectively, while $\Delta_{1} \not = \Delta$. A frequency shift occurring due to nonlinear effects existing 
in the dispersive limit applied here (see, also, \cite{zb1,sc_zb,chk,gxl}) is responsible for the resonance around $\omega$ and not exactly at this value. 
Furthermore, the environment temperature that explicitly influences the vibrational degrees of freedom, implicitly affects also the mean photon number 
in the optical resonator mode (compare solid and dashed curves) demonstrating, thus, the existence of correlations among them. The quantum nature 
of these correlations can be demonstrated via the violation of the CSI given by Eq.~(\ref{csi}). Therefore, the steady-state behaviors of the CSI versus 
$\Delta_{1}/\gamma$ is shown in Fig.~(\ref{fig-2}). The CSI is violated, i.e. ${\rm CSI} <1$, near $\Delta_{1} \approx \omega$ where the mean values 
of the photon and phonon numbers are maximal. Notice that quantum features of these correlations will disappear for $|\Delta_{1} - \omega| \gg \gamma$. 
This can be seen also from the master equation (\ref{Qq}) where the terms responsible for the cross-correlations among the two different modes will 
simple vanish in this regime, i.e. when $|\Delta_{1} - \omega|/\gamma \gg 1$.

\section{Summary}
In summary, we have studied the quantum nature of induced correlations among mechanical and optical degrees of freedom. 
A two-level quantum dot leads to correlated vibrations when fixed on a nano-mechanical resonator beam while interacting with an 
external coherent laser field as well as with an optical mode of a leaking resonator. When the variables describing the pumped 
quantum dot are faster than those of other involved subsystems and the detuning of the laser frequency from the cavity one 
is located around the vibrating mode frequency, we have found quantum correlations between the photon and phonon subsystems.  
Actually, stronger qubit-cavity coupling strengths are required that can be achieved also via involving more two-level artificial qubits. 
\acknowledgments
We are grateful to financial support via the research Grant No. 13.820.05.07/GF.



\begin{thebibliography}{33}
\bibitem{prm} X. Huang, C. Zorman, M. Mehregany, M. Roukes, Nature {\bf 421}, 496 (2003).

\bibitem{rew_nm} Y. Greenberg, Y. Pashkin, E. Il'ichev, Physics-Uspekhi {\bf 55}, 382 (2012).

\bibitem{qb1} A. Cleland, M. Geller, Phys. Rev. Lett. {\bf 93}, 070501 (2004).

\bibitem{sq1} A. Vinante, P. Falferi, Phys. Rev. Lett.  {\bf 111}, 207203 (2013).

\bibitem{exp_m} A. Safavi-Naeini, J. Chan, J. Hill, T.  Mayer Alegre, A. Krause, O. Painter,
Phys. Rev. Lett.  {\bf 108}, 033602 (2012).

\bibitem{man1} H. Okamoto, A. Gourgout, C. Chang, K. Onomitsu, I. Mahboob, E. Chang, H. Yamaguchi,
Nature Physics {\bf 9}, 480 (2013).

\bibitem{ent1} K. Brown,  C. Ospelkaus,  Y. Colombe,  A. C. Wilson, D. Leibfried,  D. Wineland,
Nature {\bf 471}, 196 (2011).

\bibitem{inter} J. Bochmann, A. Vainsencher, D. Awschalom, A. Cleland, Nature Physics {\bf 9}, 712 (2013).

\bibitem{mey} F. Bariani, J. Otterbach, H. Tan, P. Meystre, Phys. Rev. A {\bf 89}, 011801(R) (2014).

\bibitem{ent2} X.-z. Yuan, Phys. Rev. A {\bf 88}, 052317 (2013).

\bibitem{las1} I. Mahboob,  K. Nishiguchi, A. Fujiwara, H. Yamaguchi, Phys. Rev. Lett. {\bf 110}, 127202 (2013).

\bibitem{prm1} I. W. Rae, P. Zoller, A. Imamo${\rm \tilde g}$lu, Phys. Rev. Lett. {\bf 92}, 075507 (2004).

\bibitem{kmek} M. Kiffner, M. Macovei, J. Evers, C. H. Keitel, Prog. Opt. {\bf 55}, 85 (2010).

\bibitem{book} J. Peng, G.-x. Li, {\it Introduction to Modern Quantum Optics} (World Scientific, 1998).

\bibitem{DM} S. Das, M. A. Macovei, Phys. Rev. B {\bf 88}, 125306 (2013).

\bibitem{zb1} W. Ge, M. Al-Amri, H. Nha, M. S. Zubairy, Phys. Rev. A {\bf 88}, 022338 (2013); Phys. Rev. A {\bf 88}, 052301 (2013).

\bibitem{sc_zb} H. Xiong, M. O. Scully, M. S. Zubairy, Phys. Rev. Lett. {\bf 94}, 023601 (2005); 
H. T. Tan, S. Y. Zhu, M. S. Zubairy, Phys. Rev. A {\bf 72}, 022305 (2005).

\bibitem{chk} M. Kiffner, M. S. Zubairy, J. Evers, C. H. Keitel, Phys. Rev. A {\bf 75}, 033816 (2007).

\bibitem{gxl}Z.-h. Tang, G.-x. Li, Phys. Rev. A {\bf 84}, 063801 (2011); Z.-h. Tang, G.-x. Li, Z. Ficek,
Phys. Rev. A {\bf 82}, 063837 (2010); G.-x. Li, H.-t. Tan,  M.  Macovei, Phys. Rev. A {\bf 76}, 053827 (2007).

\bibitem{csi} J. F. Clauser, Phys. Rev. D {\bf 9}, 853 (1974); R. Loudon, Rep. Prog. Phys. {\bf 43}, 58 (1980).

\bibitem{prm2} J.-J. Li, K.-D. Zhu, Phys. Rev. B {\bf 83}, 245421 (2011).
\end{thebibliography}
\end{document}